\documentclass[preprint2]{aastex}

\shorttitle{Star formation history of the Small Magellanic Cloud} \shortauthors{Cignoni et al.}

\begin{document}

\title{History and modes of star formation in the most active region
  of the Small Magellanic Cloud, NGC~346\altaffilmark{1}}

\author{M. Cignoni\altaffilmark{2,3}, M. Tosi\altaffilmark{3},
  E. Sabbi\altaffilmark{4}, A.  Nota\altaffilmark{4,5}, J.S.
  Gallagher\altaffilmark{6}}

\altaffiltext{1}{ Based on observations with the NASA/ESA Hubble Space
  Telescope, obtained at the Space Telescope Science Institute, which
  is operated by the Association of Universities for Research in
  Astronomy (AURA), Inc., under NASA contract NAS5-26555. These
  observations are associated with program GO10248.}

\altaffiltext{2}{Dipartimento di Astronomia, Universit\`a degli Studi di
  Bologna, via Ranzani 1, I-40127 Bologna, Italy}

\altaffiltext{3}{Istituto Nazionale di Astrofisica, Osservatorio
  Astronomico di Bologna, Via Ranzani 1, I-40127 Bologna, Italy}
\altaffiltext{4}{Space Telescope Science Institute, 3700 San Martin
  Drive, Baltimore, USA} \altaffiltext{5}{European Space Agency,
  Research and Scientific Support Department, Baltimore, USA}
\altaffiltext{6}{Department of Astronomy, 475
  N. Charter St., Madison, WI 53706 USA}

\begin{abstract}
We discuss the star formation history of the SMC region NGC~346 based
on Hubble Space Telescope images. The region contains both field stars
and cluster members. Using a classical synthetic CMD procedure applied
to the field around NGC~346 we find that there the star formation pace
has been rising from a quite low rate 13 Gyr ago to $\approx 1.4\times
10^{-8}\,M_{\odot}\,yr^{-1}pc^{-2}$ in the last 100 Myr. This value is
significantly higher than in other star forming regions of the
SMC. For NGC~346 itself, we compare theoretical and observed
Color-Magnitude Diagrams (CMDs) of several stellar sub-clusters
identified in the region, and we derive their basic evolution
parameters. We find that NGC~346 experienced different star formation
regimes, including a dominant and focused ``high density mode'', with
the sub-clusters hosting both pre-main sequence (PMS) and upper main
sequence (UMS) stars, and a diffuse ``low density mode'', as indicated
by the presence of low-mass PMS sub-clusters. Quantitatively, the star
formation in the oldest sub-clusters started about 6 Myr ago with
remarkable synchronization, it continued at high rate (up to $2\times
10^{-5}\,M_{\odot}\,yr^{-1}\,pc^{-2}$) for about 3 Myr and is now
progressing at a lower rate. Interestingly, sub-clusters mainly
composed by low mass PMS stars seem to experience now the first
episode of star formation, following multi-seeded spatial patterns
instead of resulting from a coherent trigger. Two speculative
scenarios are put forth to explain the deficiency of UMS stars: the
first invokes under-threshold conditions of the parent gas; the second
speculates that the initial mass function (IMF) is a function of time,
with the youngest sub-clusters not having had sufficient time to form
more massive stars.

\end{abstract}


\section{Introduction}

The Small Magellanic Cloud (SMC) is the closest star forming galaxy
with a low metallicity ($Z=0.004$) typical of late-type dwarfs and
most similar to that of primordial galaxies. For this reason an
increasing number of studies are being devoted to its star formation
history (SFH) and related processes. This research is part of a
long-term project aimed at studying how the star formation started and
propagated in the SMC, studying both young clusters and the field
population. We concentrate here on the OB association NGC~346, the
most active star forming region, where large numbers of pre-main
sequence and massive stars are found. The inherent complexity of this
star forming cloud is well recognized. This region provides an
excellent sample of newly formed stars spanning a wide range of
masses, and bridging a wide range of temporal and spatial scales.

 The results presented in this paper are derived from observations
 acquired with the Advanced Camera for Survey (ACS) on board of the
 HST. Several photometric investigations tackled this complex
 population; all of them, using isochrone fitting, agreed that the
 star formation has taken place in a variety of sub-clusters at
 different local conditions. However, different formation scenarios
 are proposed: \cite{sabbi2007} suggested a nearly coeval star
 formation in the cloud about 3 Myr ago. \cite{contursi2000} proposed
 a progressive star formation from the central cluster and propagating
 along the Bar. \cite{gouli2008} argued that NGC~346 had been shaped
 by two delayed triggering events, the first one due to the central
 cluster and the second one to the massive progenitor of
 SNR~B0057-724.

The purpose of this study is to 1) re-examine the star formation of
the individual star-forming sites to a much finer spatial scale
($\sim\,1$ pc), by means of the synthetic CMD approach, and 2) evaluate
systematically whether and how the IMF and the star formation rate
(SFR) are shaped and modified by the environment.

\section {Stellar content and its spatial distribution in the region of NGC 346}

\begin{figure}[t!]
\centering \includegraphics[width=8cm]{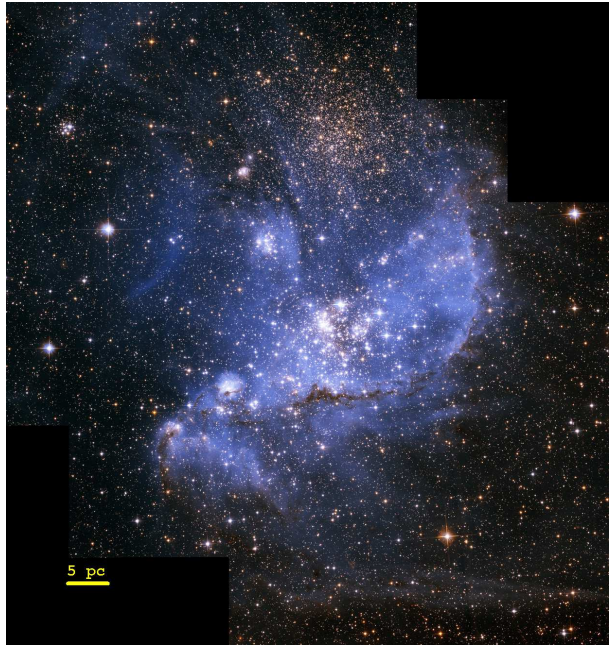}
\caption{ACS image for the observed region around NGC~346
  (\citealt{nota2006}).}
\label{hst} 
\end{figure}

Fig. \ref{hst} shows our HST image of NCG~346, acquired with three
overlapping pointings of the ACS Wide Field Channel
(\citealt{nota2006}). The data analysis for this data set is presented
in \cite{sabbi2007} (hereafter S07). Assuming an intrinsic distance
modulus to the SMC of $(m-M)_0=18.9$, the field of view covers about
$88 \times 88\, pc^2$ and contains different stellar populations, both
clustered and diffuse. The clustered population has two main
components: the intermediate age cluster BS~90 (\citealt{bica1995}),
clearly recognizable as a roundish system at the top (north) of the
image, and several clumps of stars (``sub-clusters'' according to the
definition by S07), spread over the field along a sort of umbrella
shape, and identified by S07.  The diffuse population is uniformly
spread over the field.
BS~90 is 4-5 Gyr old and most likely located at the foreground of the
NGC~346 cluster (see S07 and \citealt{cignoni2010}). The sub-clusters
are all part of the NGC~346 region, with ages from a few to 20 Myr
(S07, \citealt{gouli2008}, \citealt{cignoni2010}), while the diffuse
population is presumably representative of the field SMC population in
that region. The latter is then composed by SMC fore and background
stars along the line-of-sight, possibly covering several kpc (see
e.g. \citealt{glatt2008}), and spanning a wide range of ages, from Myr
to ten Gyr.

Our goal is to study the SFH of all these components by interpreting
their observational CMDs with synthetic ones. BS~90 has already been
studied by S07 (see also \citealt{rochau2007}) and no further analysis
is presented here. We thus concentrate on the other two
components. The derivation of the SFH of the diffuse component is a
standard application of the synthetic CMD method (see, e.g.,
\citealt{tosi1991}, \citealt{cignoni2006}, \citealt{cignoni2010b}) to
a statistically significant sample of field stars properly located in
the region far from the sub-clusters. On the other hand, the analysis
of the clustered component is handled with a dedicated methodology for
two main reasons: 1) the various sub-clusters appear to have somewhat
different ages and therefore we cannot treat them all together; 2)
each sub-cluster contains few stars and therefore, when treated
individually, has rather large associated statistical
uncertainties. So, for the sub-clusters we have compared the observed
CMDs with each other and with synthetic ones, taking into account the
low number statistics and systematics.

Following the classification scheme proposed in \cite{contursi2000},
the NGC~346 region can be divided in a Spur, a filamentary low density
structure oriented to the North-East direction, and a fan-shaped
feature (Bar), hosting the bulk of intermediate and massive stars.

 The first question is whether two stellar tracers of the most recent
 activity, namely objects on the UMS and objects still on the PMS, are
 sharing the same ``fine'' spatial structure. A similar analysis has
 been performed by \cite{schmeja2009} with the goal to derive the
 clustering behavior, while the purpose here is to identify
 sub-clusters with peculiar mass function and perform an accurate CMD
 analysis.

The main advantage in using two different tracers stems from their
different observational and intrinsic uncertainties. With respect to
UMS stars, PMS stars suffer less external contamination and low number
statistics. On the other hand, UMS stars are less affected by
incompleteness and theoretical uncertainties.

\begin{figure*}[t!]
\centering\includegraphics[width=13cm]{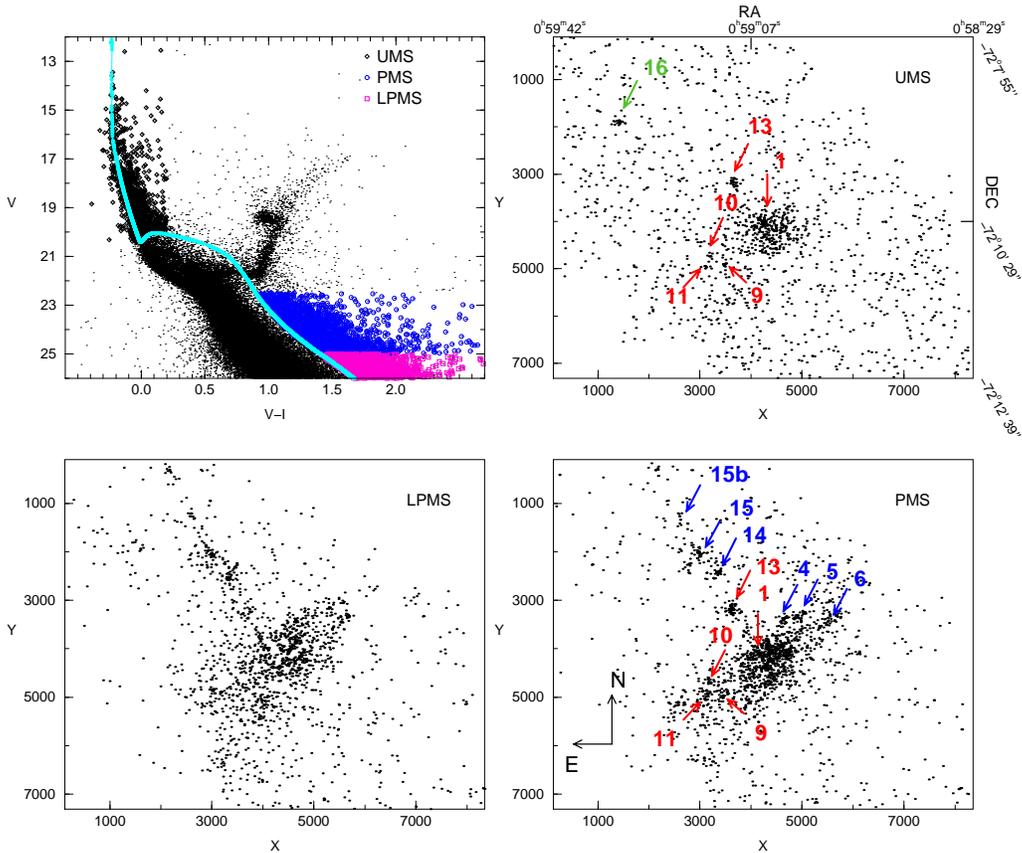}
\caption{Top-left panel: CMD for the region NGC~346. Black open
  diamonds, blue open circles and magenta open squares indicate the
  areas referred to the UMS, PMS and LPMS stars, respectively. The
  line corresponds to a 3 Myr isochrone including the PMS phase
  (\citealt{cignoni2009}). The other panels (from top-right clockwise)
  show the corresponding spatial distributions. Red arrows indicate
  sub-clusters hosting UMS and PMS stars; blue arrows refer to PMS
  dominated sub-clusters; and the green arrow refers to the only UMS
  sub-cluster. The number labeling each sub-cluster follows the
  nomenclature provided by \cite{sabbi2007}, where the whole catalogue
  with the star coordinates can also be found. In the top-right panel
  we shows also the RA and DEC coordinates of the plot extremes and of
  its center (X=Y=4000). }
\label{upms_pms} 
\end{figure*}

The regions of the CMD we associate with UMS and PMS stars are
indicated in the top-left panel of Fig. \ref{upms_pms} (with different
colors in the electronic version).
 For UMS stars (black open diamonds) we chose all objects above the
 Turn-On (TOn\footnote{The TOn is the point in the color-magnitude
   diagram where the PMS stars join the main sequence (see
   e.g. \citealt{cignoni2010}).}) of a 3 Myr isochrone and bluer than
 $V-I=0.2$. With this definition, the UMS sample is mostly composed by
 intermediate mass stars and by a few massive stars. For the PMS stars
 we considered two samples: objects redder than the 3 Myr isochrone
 with magnitude in the range $22.5<V<25$ (blue open circles,
 hereinafter PMS sample) and fainter than $V=25$ (magenta open
 squares, low mass PMS sample, hereinafter, LPMS). It is easy to
 notice that while the PMS sample is younger than 3 Myr (or only
 slightly older if a modest additional reddening is taken into
 account), the UMS sample can include MS stars as old as 600
 Myr. Moreover, while there is no doubt that the PMS and LPMS samples
 are free from any contamination, the UMS sample can include a minor
 fraction of PMS stars starting to approach the MS.

The top-right, bottom right and bottom left panels of Figure
\ref{upms_pms} show the location of the selected UMS, PMS and LPMS
stars, respectively. These distributions provide a clue on the history
of the region: half of the UMS stars are clumped into a \emph{few
  agglomerates} (SC-1, SC-13 and SC-16, named according to S07), while
the other half is more evenly distributed, and are probably members of
the SMC field (foreground and background stars). One of the most
striking aspects of this distribution is the absence of filamentary
structures and a rather round appearance of the sub-clusters.

In contrast, the PMS stars are found almost exclusively either clumped
or irregularly arranged along filaments. More filamentary than clumpy
is also the distribution of the LPMS sample. In particular, we note
that the Spur region, composed by a few aggregates in the PMS map,
becomes a sort of bridge extending for tens of pc in the LPMS map. A
word of caution is however necessary for the frequent holes in their
star distribution: like that clearly visible in the center of SC-1,
they are likely due to incompleteness effects. Incompleteness is
caused by crowding, which reaches a maximum in the central region
SC-1, and is much less severe at the outskirts of the region.

\section {Sub-cluster properties}
Looking at the maps of Figure \ref{upms_pms}, there is a further
intriguing aspect of the star spatial distribution: for reasons that
will be discussed later, not all the sub-clusters visible in the UMS
map are detected in the PMS map, and vice versa. In Figure
\ref{upms_pms} stellar sub-clusters which host both species are
indicated with red arrows, while blue and green arrows indicate
sub-clusters which host only PMS or UMS stars, respectively. We find
that the PMS sub-clusters reside in the North-East part of the Spur
(NE-Spur) (SC-14, SC-15 and SC-15b) and in the North-West side of the
Bar (NW-Bar) (SC-4, SC-5 and SC-6). On the other hand, UMS-PMS
sub-clusters are found in the central region (SC-1), in the southern
cluster of the Spur (SC-13) and in the South-East side of the Bar
(SE-Bar) (SC-9, SC-10 and SC-11). The only sub-cluster to host
exclusively UMS stars is SC-16. In fact, SC-16 is older than 10 Myr
(see e.g. S07 and \citealt{cignoni2010}), and its PMS members are too
blue to be included in our PMS selection. The situation is radically
different for SC-14, SC-15 and SC-15b.  Despite the few Myr age of
these sub-clusters, there are few, if any, UMS stars: where is the MS
counterpart of their observed PMS stars? 

\begin{figure*}[t]
\centering \includegraphics[width=8cm]{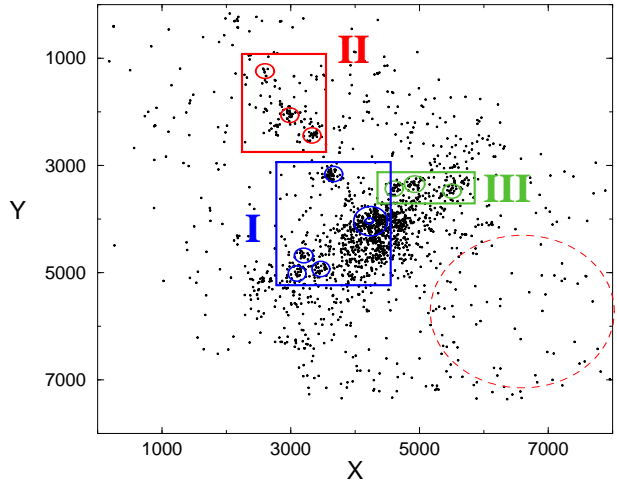}
\caption{The three boxes superimposed to the NGC346 map identify
  sub-clusters with similar properties. Solid circles indicate the
  sub-clusters whose CMDs are discussed in the text. The dashed large
  circle represents the region of field that has been used to estimate
  the contamination (see text).}
\label{cerchi} 
\end{figure*}

\begin{figure*}[t]
\centering\includegraphics[width=16.0cm,angle=0]{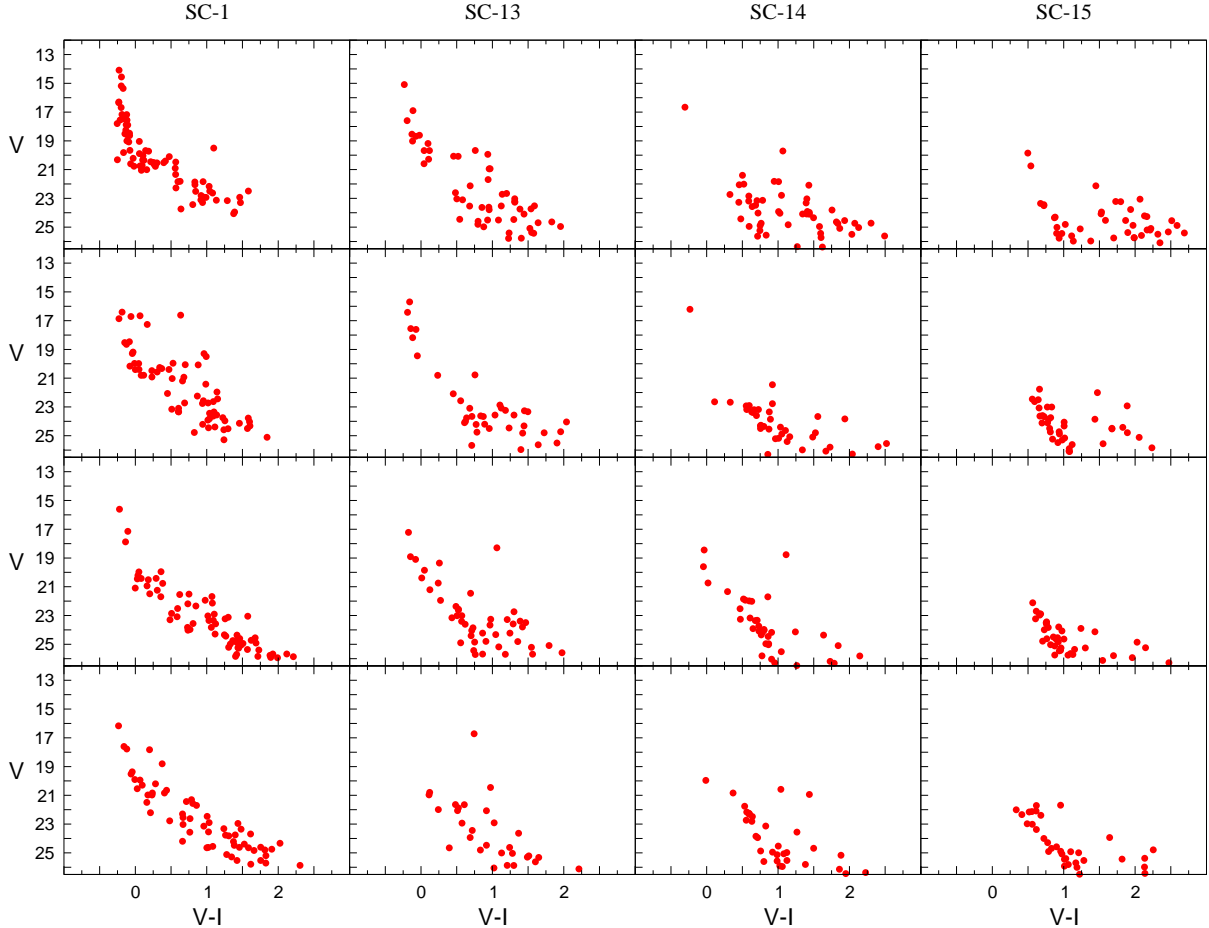}
\caption{From left to right, the four columns show CMDs for SC-1,
  SC-13, SC-14 and SC-15 respectively. From top to bottom, stars are
  shown that belong to progressively more external annuli of equal
  area centered on the highest density peak. The radius of the inner
  circle is about 71 pixels ($\approx\,1$ pc) for all sub-clusters. }
\label{cmds_radial} 
\end{figure*}

To try and understand the sub-cluster similarities and differences, we
have divided them in three categories: Group I sub-clusters are those
with minimum PMS/UMS ratio, Group II sub-clusters those with highest
PMS/UMS ratio, while we have labelled as Group III the sub-clusters
where the ratio was either intermediate or difficult to
estimate. Fig. \ref{cerchi} shows the clear spatial separation of the
three different Groups.
Figure \ref{cmds_radial} shows the CMD for typical Group I sub-clusters
(SC-1 and SC-13) and for Group II sub-clusters (SC-14 and SC-15) as a
function of the distance from the sub-cluster center. Different rows
refer to stars selected from four equal area annuli.

Excluding the very central region of SC-1, affected by incompleteness,
these CMDs trigger two interesting questions:

1) {\emph {Lack of UMS stars}}: As suggested by the spatial maps,
intermediate and massive MS stars seem deficient in sub-clusters like
SC-14.  Massive stars are always rare and sub-clusters like SC-14 are
tiny pockets of stars; hence, stochastic fluctuations due to small
number statistics in the poorly populated high mass end of the IMF
need to be carefully accounted for. Is this sufficient to conclude
that the star formation in SC-14/15 is a simple scaled-down version of
that of SC-1?

2) {\emph {Redness of PMS stars}}: The red tail of the PMS distribution
extends to redder colors going from SC-1 to SC-15 along the spur. Is
their redness intrinsic (age dispersion or circumstellar disks) or
caused by obscuring material along the line of sight?

From the analysis of the sub-clusters CMD we find that the Group I
members show minimum PMS redness, minimum UMS spread and, by
definition, minimum PMS/UMS ratio. The Group II members show the
maximum PMS redness and, by definition, the highest PMS/UMS ratio. The
Group III members resemble Group I members in relation to the PMS
redness and to Group II in relation to the PMS/UMS ratio, while the
large color spread among UMS stars is a unique property of this class
of sub-clusters.

The next section is dedicated to a quantitative analysis of these
groups. In order to reproduce the observed CMD morphology and star
counts of the sub-clusters shown in Figure \ref{upms_pms} we applied
the synthetic CMD approach, combining the $Z = 0.004$ Pisa stellar
models for PMS stars (see \citealt{cignoni2009}) with the same
metallicity Padova stellar models (\citealt{fagotto1994}) for later
evolutionary phases. The explored mass range is between $0.45$ and
$120\,M_{\odot}$. To produce realistic simulations we have
incorporated photometric errors and incompleteness corrections as
derived from extensive artificial tests on the real images
(S07). Different parameters, such as extinction, IMF, binarity, and
star formation rate are varied until the stellar densities and
distributions well match the observed CMDs.

To better interpret the morphology of the sub-cluster sequences we
chose sub-cluster radii as the best compromise between the need to
bypass the crowding problems, typical of the very inner central
regions, and the need to minimize the effect of field star
contamination, while still containing a reasonable number of
sub-cluster stars. Concerning SC-1, we find that the best strategy is
to focus on an annulus around the center from 50 to 282 pixels (see
Figure \ref{cerchi}). Artificial star tests indicate that in this area
stellar detections are more than 90\% complete down to $V\approx
24$. For all minor and satellite sub-clusters we find that a circle of
radius 141 pixels (see Figure \ref{cerchi}) allows to include most of
the members with a modest field contamination.

\section{Group I}

\subsection{SC-1}

\begin{figure}[t!]
\centering \includegraphics[width=8cm]{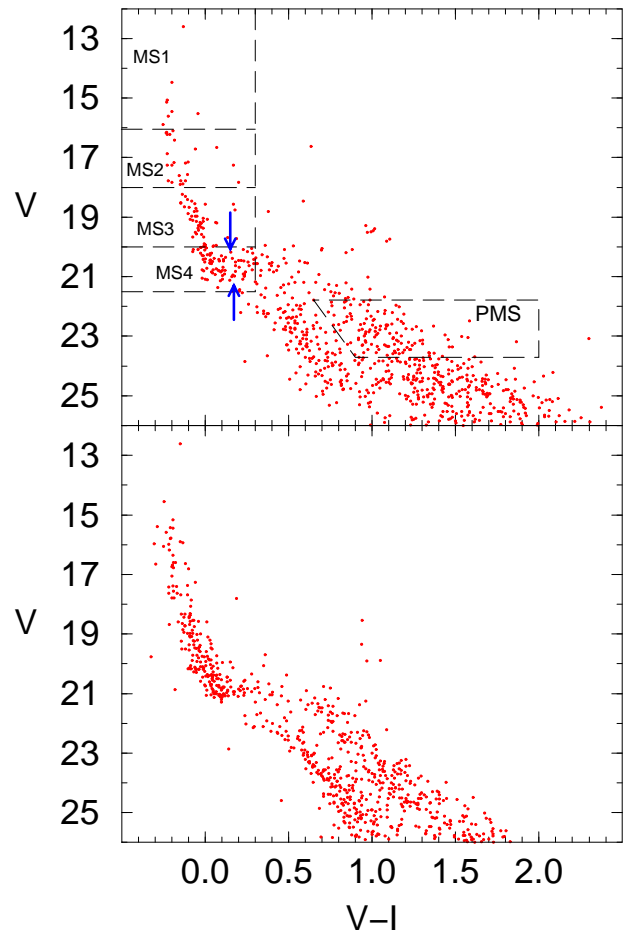}
\caption{Comparison of observational (top panel) and best fit
  synthetic CMD (bottom panel) for the sub-cluster SC-1. Also shown
  are the regions used for characteristic star counts (see
  text). There are four such regions along the MS and one for the PMS
  phase. The theoretical CMDs are calculated with a 30\% of binaries
  and Salpeter IMF.  In the field shown in the figure, the
  observational and theoretical CMDs contain the same number of
  stars.}
\label{sc1} 
\end{figure}

SC-1 is the central and most populous sub-cluster in NGC~346. We have
simulated synthetic CMDs based on different combinations of
evolutionary (SFH, IMF) and environmental parameters (reddening,
distance modulus, fraction of binary stars). The best combination of
parameters is assessed by examining the star counts in strategic boxes
along the main sequence (see labels MS1 to MS4 in Figure \ref{sc1},
top panel) and in the pre-main sequence (PMS). Given the youth of this
sub-cluster, likely younger than 10 Myr, regions MS1 and MS2 are
particularly suitable to constrain the IMF. MS3 and MS4 convey
information both on the IMF and on the star formation history. The PMS
region informs mainly on the IMF (given the uncertainties on the PMS
models, the PMS box is used only as, a posteriori, consistency
check). From a numerical point of view, a grid search routine is used
to effectively determine the combination of parameters minimizing
residuals in these regions. To limit the parameter space we also
considered two additional morphological features: the magnitude V
dispersion of the PMS stars at the TOn (marked with arrows in the top
panel of Figure \ref{sc1}), and the number of MS stars in the range of
magnitude 21.3 and 21.8 (approximately the 6 Myr TOn and the BS~90
Turn-Off), which is a strong indicator of contamination and/or
sub-cluster members older than about 6 Myr. Our solution to handle
this contamination was to use as control fields hundred regions of
equal area located in the south-west part of the image (see the large
dashed circle in Figure \ref{cerchi}).

 As a first result, a Salpeter IMF (\citealt{salpeter1955}), a
 fraction of binaries of 30\%, a foreground reddening $E(B-V)=0.08$
 and a distance modulus ${(m-M)}_{0}=18.9$ provide a good agreement
 with the data. Concerning the SFH, we find that the stellar
 production in SC-1 started energetically between 5 and 6 Myr ago, was
 strong for about 3 Myr (period between 3 and 6 Myr ago), then it
 dropped, probably victim of feedback from the massive stars of the
 first generation which quenched the subsequent formation. Before the
 onset at 6 Myr, the observed counts in the CMD window between the 6
 Myr TOn and the BS~90 Turn-Off suggest a null or negligible activity:
 out of 9 objects in the magnitude interval $21.3\,-\,21.8$ and
 $V-I<0.45$ at least three can be attributed to the field with a
 confidence level better than 95\%. The residual stars are compatible
 with a star forming activity of at most $0.3\times
 10^{-5}\,M_{\odot}\,yr^{-1}\,pc^{-2}$ in the period between 8 Myr and
 6 Myr ago. During the active phase this region experienced a
 peak\footnote{Actually, this is a lower limit, which does not take
   into account stars below $0.45\,M_{\odot}$. When extrapolated using
   a Salpeter slope down to $0.1\,M_{\odot}$, this translates in an
   upper limit of $3.6\times 10^{-5}\,M_{\odot}\,yr^{-1}\,pc^{-2}$.}
 of about $2\times 10^{-5}\,M_{\odot}\,yr^{-1}\,pc^{-2}$ between 4 and
 5 Myr ago, a value much higher than in the center of NGC~602 (see
 \citealt{cignoni2009}), another active star forming region in the
 SMC. After this onset, the MS boxes strongly exclude that the star
 formation was constant or increasing to the present day: according to
 our best model, only 20\% of the total mass of young stars in the
 explored region is produced in the last 3 Myr. Moreover, the
 smoothness of the UMS is not suggestive of any recent short burst.

Figure \ref{sc1} shows a comparison between our best synthetic CMD
(bottom) with the observational one (top).
For better visualization, the model has been artificially
``contaminated'' with field stars taken from an appropriate region of
equal area. In broad terms, the UMS and the TOn regions are well
reproduced, although the synthetic main sequence appears more
fuzzy. Moreover, two discrepancies are noted: 1) our models do not
provide satisfactory results about the dispersion of PMS stars,
predicting smaller spreads and bluer colors than observed (see also
\citealt{pozzo2003} and \citealt{mayne2007}); 2) no combination of
parameters reproduces the observed number of PMS stars. Once the MS
boxes are matched, our best model underestimates it systematically by
about 40\%. Although part of this difference may arise from our lack
of PMS models less massive than $0.45\,M_{\odot}$ (potentially present
in the PMS box for ages younger than few hundreds of Kyr), a complete
solution of the discrepancy seems to require other physical
reasons. It is intriguing to note that such discrepancy could be
easily resolved by considering an additional episode of star formation
younger than 2 Myr, accounting for about $500\,M_{\odot}$, where the
MS phase is artificially suppressed and, therefore, not visible in the
CMD. As an alternative, the IMF could be steeper than Salpeter's for
masses below $2\,M_{\odot}$, thereby creating a larger population of
low mass stars.

In the following we adopt SC-1 as reference sub-cluster to compare
with minor sub-clusters.

\subsection{SC-13}

\begin{figure}[t!]
\centering\includegraphics[width=8cm]{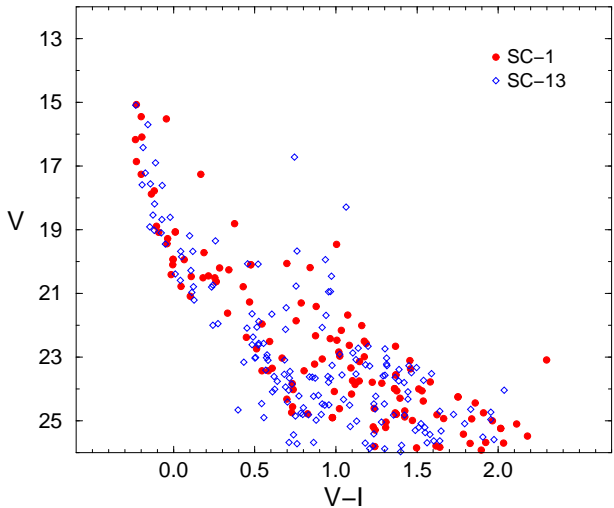}
\caption{Comparison between SC-1 (filled red circles) and SC-13 (open
  blue diamonds). The CMD for SC-1 is normalized to the number of
  SC-13 stars brighter than $V=22$.}
\label{sc1_sc13} 
\end{figure}

The SC-13 sub-cluster is less dense than SC-1: excluding MS stars
fainter than $V=22$, it accounts for about 7 stars per $pc^2$ against
about 11 stars per $pc^2$ found in SC-1. Moreover, SC-13 is less
affected by incompleteness. Despite these differences, when the CMD of
SC-1 is normalized to the number of SC-13 stars brighter than $V=22$
(inside a radius of 141 pixels), the CMD morphologies appear very
similar (see Figure \ref{sc1_sc13}).
First, the color dispersion along the upper main sequence and the TOn
magnitude are indistinguishable from one another, suggesting an
equivalent star formation duration and onset. Second, the PMS region
displays an identical spread and the ratio PMS/UMS is fairly similar
(about 2, see Table \ref{tab}).
\begin{table}[h!]
  \caption{UMS, PMS, LPMS star counts and PMS/UMS ratio (P/U) for the
    labeled sub-clusters. All counts are measured inside a radius of
    141 pixels, apart SC-1 that is measured in an annulus around the
    center between 50 and 282 pixels.}
  \label{tab}
  \begin{center}
    
    \begin{tabular}{|r|l|l|l|l|l|}
      \hline
      & & UMS&PMS&LPMS&P/U \\\cline{1-6}
      & SC-1 & 94&195&79&2.1 \\\cline{2-6}
      Group I & SC-13&20&35&13&1.8 \\ \cline{2-6}
      & SC-9 &10&19&7& \\\cline{2-5}
      & SC-10&7&19&1& 2.6 \\\cline{2-5}
      & SC-11&7&24&18& \\\cline{2-5}
      \hline \hline
      & SC-14 & 5&23&19& \\\cline{2-5}
      Group II & SC-15&0&24&24&10.8 \\ \cline{2-5}
      & SC-15b &0&7&11& \\\cline{2-5}
      \hline \hline
      & SC-4 & 2&22&19& \\\cline{2-5}
      Group III & SC-5&2&29&11&10.4 \\ \cline{2-5}
      & SC-6 &3&22&18& \\\cline{2-5}
      \hline
    \end{tabular}
  \end{center}
\end{table}
As already envisaged in \cite{cignoni2010}, the TOn gets brighter (up
to 1 mag) when stars from the inner region of SC-13 (see the top row,
second column of Figure \ref{cmds_radial}) are selected. Although
numbers are too small to allow statistically significant conclusions,
these findings suggest that: 1) the star formation in SC-13 was
triggered at the same time of SC-1; 2) the inner region of SC-13 may
have experienced a secondary star forming episode about 3 Myr ago. In
terms of rate density, we estimate that SC-13 astrated (at peak) about
$1.3\times 10^{-5}\,M_{\odot}\,yr^{-1}\,pc^{-2}$.

\subsection{SC-9, SC-10, SC-11}

\begin{figure}[t!]
\centering
\includegraphics[width=8cm]{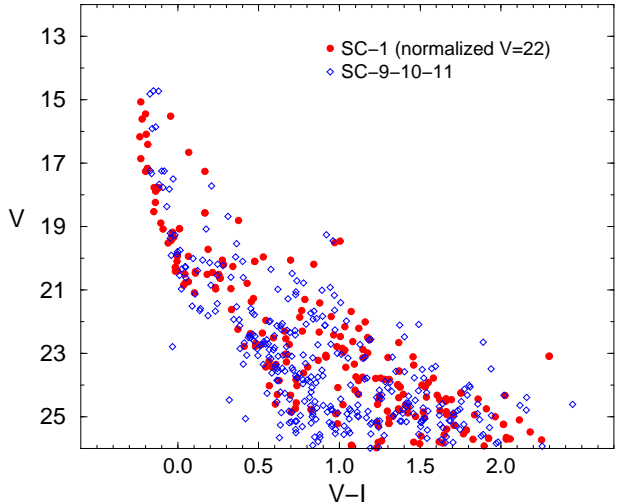}
\caption{Composite CMD for SC-9, SC-10 and SC-11 (open blue diamonds)
  superimposed to the CMD for SC-1 (filled red circles) normalized to
  have the same number of stars brighter than $V=22$. }
\label{sc10} 
\end{figure}

In Figure \ref{sc10} we present (blue open diamonds) the combination
of the CMDs of the sub-clusters SC-9, SC-10 and SC-11, all located in
the SE-Bar, superimposed to SC-1 (conveniently normalized to the same
number of stars brighter than $V=22$). The similarity of morphologies
is remarkable: UMS, intermediate mass PMS and PMS spread overlap. In
terms of star counts, the ratio PMS/UMS (about 2.6, see Table
\ref{tab}) is consistent within the expected Poisson fluctuations.

 When normalized to the same area, we estimate the maximum star
 formation rate density in the SE-Bar to be $0.4\times
 10^{-5}\,M_{\odot}\,yr^{-1}\,pc^{-2}$ .

\section{Group II}

\begin{figure}[t!]
\centering
\includegraphics[width=8cm]{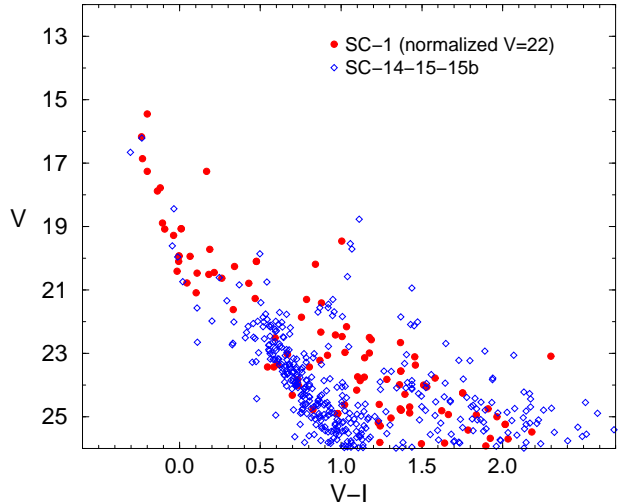}
\caption{Comparison between Group II sub-clusters SC-14, 15, 15b (open
  blue diamonds) and SC-1 (filled red circles) re-sampled to have the
  same number of stars brighter than $V=22$.}
\label{sc14} 
\end{figure}

All located in the NE-Spur, the members of this group are embedded in
dust and nebulosities. Unlike Group I, these sub-clusters are mainly
composed of low-mass stars with few, if any, UMS stars. This condition
strongly hinders any attempt to obtain reliable ages using the
TOn. Figure \ref{sc14} shows a direct comparison between the composite
CMD of SC-14, 15 and 15b with SC-1 (re-sampled to have the same number
of bright stars with $V<22$). 
Two differences emerge: 1) the Group II CMD shows a ratio PMS/UMS
stars of about 11 (see Table \ref{tab}), at least a factor of three
larger than in SC-1, and a lack of intermediate mass PMS stars, which
are clearly present in the CMD of SC-1 at $V=20$ and in the color
range $0.1<V-I< 1$; 2) In the Group II CMD the LPMS stars are much
redder than in SC-1.

Concerning point 1), we notice however that the presence of
intermediate mass and massive young stellar objects (YSO) has been
discovered with Spitzer observations by \cite{simon2007}, suggesting
that a fraction of UMS stars may exist but be still invisible in
optical wavelengths.
\begin{figure}[t!]
\centering \includegraphics[width=8cm]{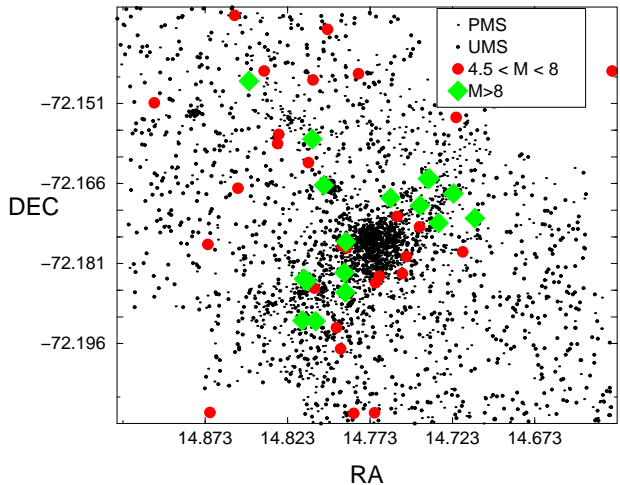}
\caption{Spatial distribution of the Spitzer sample of intermediate
  and high mass YSOs (data from \citealt{simon2007}) overlaid on the
  map of PMS (black dots) and UMS stars (large black dots) from Figure
  \ref{upms_pms}). Red circles and green diamonds stand for YSOs with
  estimated mass between 4.5 and 8 $M_{\odot}$ and larger than
  $8\,M_{\odot}$, respectively.}
\label{simon} 
\end{figure}
We show in Figure \ref{simon} the distribution of Spitzer YSOs in our
field of view. 

Point 2) is not caused by either dust between us and the sub-cluster
or diffuse dust within the sub-cluster (otherwise also the lower main
sequence would exhibit the color excess) but rather by reddening
material intimately related to the individual PMS stars. It is also
noteworthy that the UMS is not reddened: either these objects belong
to the field or the reddening material affects only the PMS phase (as
expected for circumstellar material).

\begin{figure}[t!]
\centering \includegraphics[width=8cm]{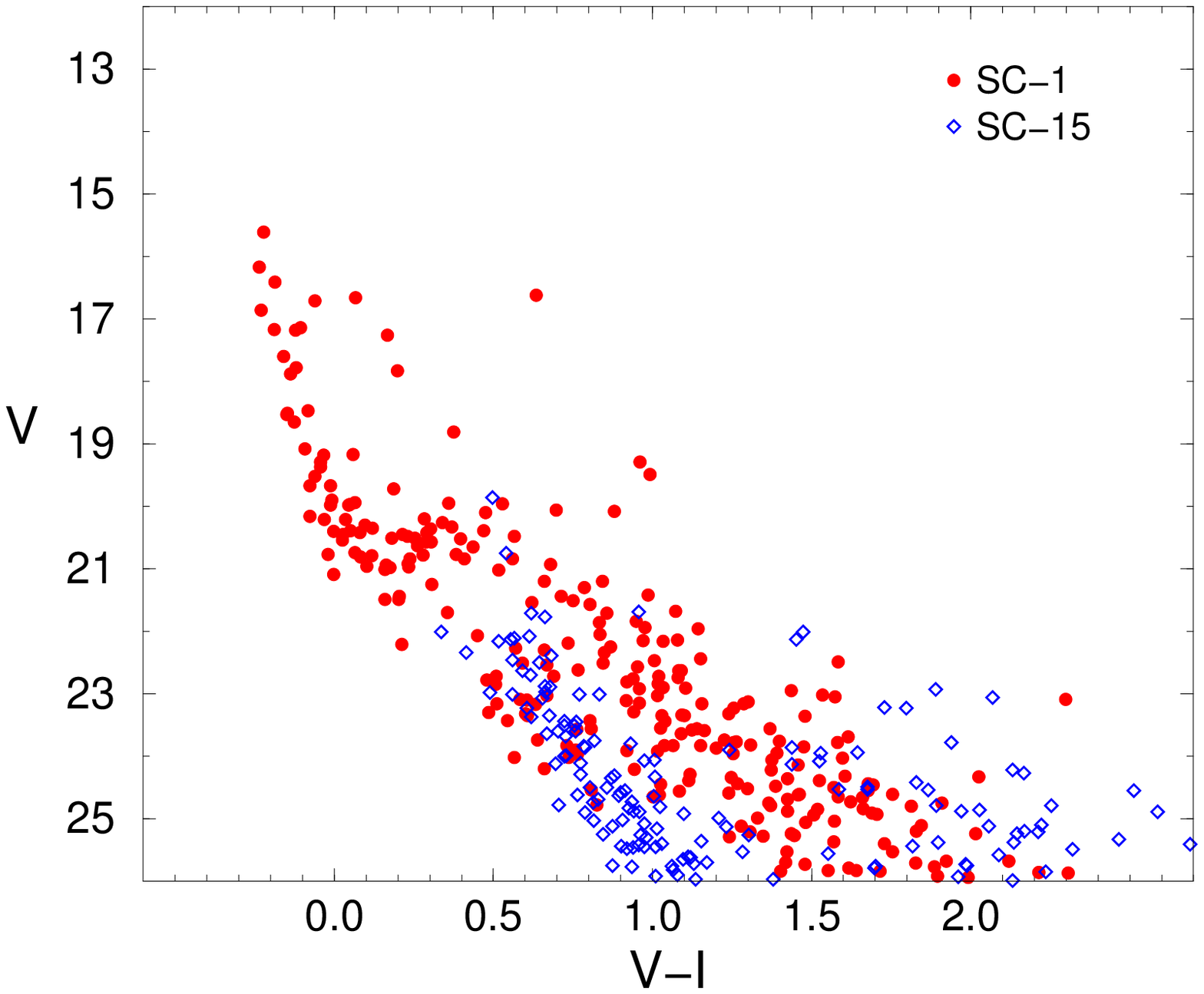}\\
\centering \includegraphics[width=8cm]{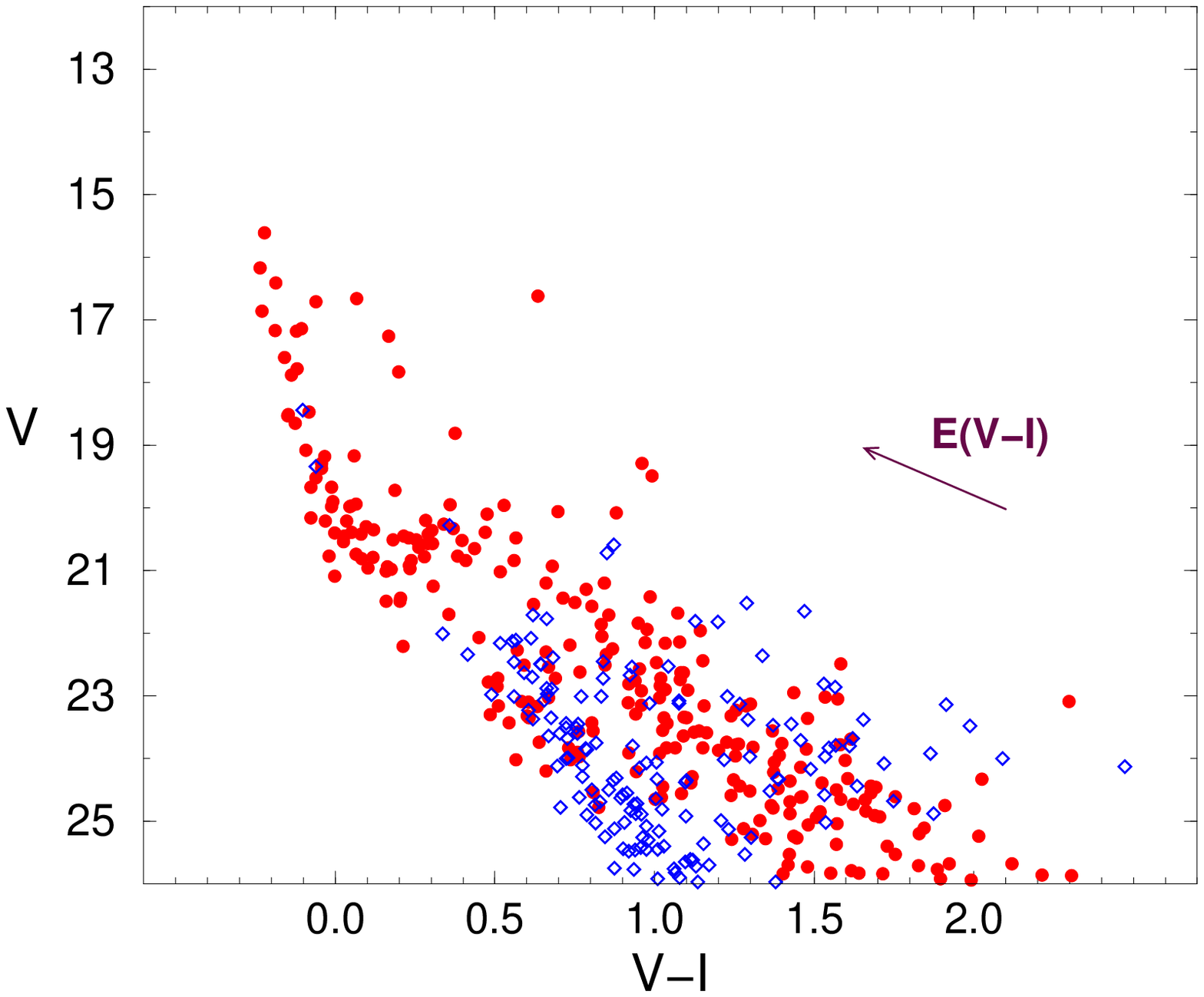}
\caption{Top panel: comparison between SC-15 (open blue diamonds) and
  SC-1 (filled red circles) normalized to the same area. Bottom panel:
  SC-15 after reddening correction applied to PMS stars only.}
\label{sc15} 
\end{figure}

Among the Group II sub-clusters, SC-15 (Figure \ref{sc15}) is the one
displaying more significant differences with respect to SC-1. The
upper main sequence is definitely underpopulated, with just a couple
of stars at best. In addition, the LPMS stars in SC-15 are at least
0.5 mag redder than comparable objects in SC-1, while the main
sequence stars fainter than $V=22$ are only slightly redder (0.1
mag). Once again, this would support the idea that reddening material
is differentially distributed among PMS stars.

In an attempt to reconcile the PMS samples, we shifted the SC-15 stars
along the reddening vector until the color distribution of PMS in the
two sub-clusters CMD overlapped. We derived the optimum shift when the
two brighter stars in the SC-15 PMS sample reached the main sequence
color. Figure \ref{sc15}-(bottom panel) shows the CMD obtained with
this methodology. It is clear that the resulting correction,
$E(V-I)\sim 0.6$, is still largely insufficient to reconcile the bulk
of PMS stars. Only a differential reddening may overcome the remaining
discrepancy. However, the reddening vector (see Figure \ref{sc15},
bottom panel) is almost parallel to the PMS and a full (differential)
correction would produce too many bright PMS stars. In other words,
although a strong extinction by circumstellar envelopes would be very
likely in such young objects, it would necessarily further increase
the ratio of PMS stars over UMS stars.

\section{Group III}

\begin{figure}[t!]
\centering \includegraphics[width=8cm]{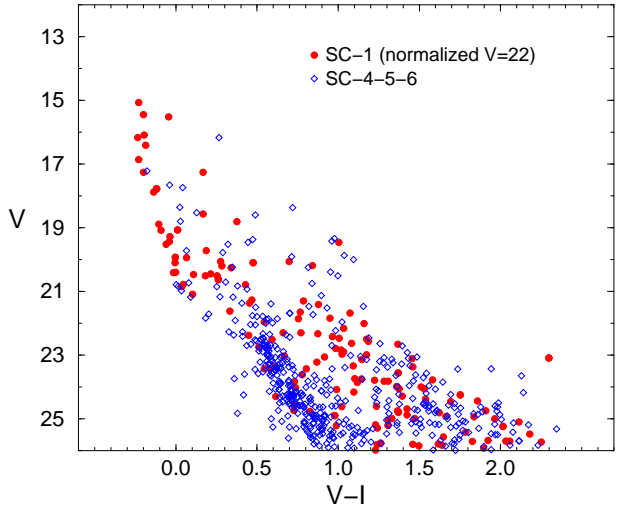}
\caption{Comparison between the composite CMD for Group III
  sub-clusters SC-4, SC-5 and SC-6 superimposed to the CMD for SC-1
  normalized to have the same number of stars brighter than $V=22$.}
\label{sc456} 
\end{figure}

All members of this group are located in the NW-Bar. Figure
\ref{sc456} shows the combination of CMDs for the sub-clusters SC-4,
SC-5 and SC-6 superimposed to the CMD for SC-1. The stellar population
in this group shows CMD features which are somehow intermediate
between Group I and Group II. The color spread in the LPMS is fully
consistent with the same sequence in SC-1. On the other hand, the
ratio PMS/UMS of about 10.4 (see Table \ref{tab}) is close to the
value found in Group II. Moreover, while in SC-1 the UMS is sharply
defined and an extended sequence of intermediate mass PMS stars is
observed, all bright ($V<23$) stars in the Group III CMD seem not to
lie along any recognizable sequence. Also for these sub-clusters
\cite{simon2007} report the detection of massive embedded YSOs (see
Fig. \ref{simon}).

Finally, it is worth noticing that the field contamination,
represented by MS stars fainter than $V=22$, is stronger in the NW-Bar
and in the NE-Spur than in the SE-Bar because the former regions are
closer to the cluster BS~90.

\section{Field Star Formation History}

\label{sfh}

In addition to a very young population, clearly reflected by clumps or
filaments mainly studded by PMS stars, Figure \ref{upms_pms} reveals
the signatures of an evenly distributed component of UMS stars without
an equivalent counterpart of PMS stars. This diffuse and ``UMS
dominated'' component is clearly inconsistent with a Myr old
population and corresponds to a (pervasive) presence of SMC field
stars.

In the previous section such field population has been considered only
as a mere intruder of the young sub-clusters. Nevertheless, field stars
also retain valuable information on the average SFH in the
region. Such history is not represented by either BS~90 or the
sub-clusters, since these structures are distinct snapshots in space and
time (Gyrs old the former, Myrs old the latter) of the overall history
in the SMC.

\begin{figure}[t!]
\centering \includegraphics[width=8cm]{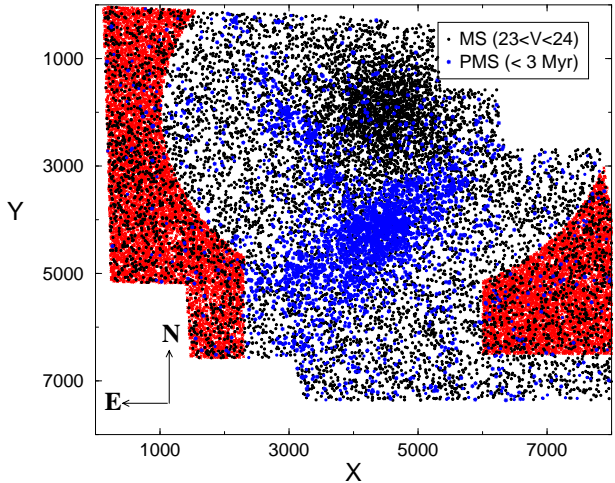}
\caption{Spatial position of the selected field stars (red dots), PMS
  stars younger than 3 Myr (blue dots) and main sequence stars with
  $23<V<24$ (black dots; see Fig.3 in \citealt{cignoni2010}),
  respectively. }
\label{xy_campo} 
\end{figure}

We present here a quantitative analysis of the field SFH. To this aim,
field stars (red dots in Fig. \ref{xy_campo}) have been chosen as
isolated as possible to avoid contamination from either BS~90 or the
sub-clusters.  However, given the radial profile of BS~90 (see S07), the
lower main sequence (fainter than $V\approx 22$), as well as the red
giant branch (RGB) and the red clump (RC) may still suffer of a
residual contamination, potentially leading to an overestimate of the
field star formation between 4 and 5 Gyr.

\begin{figure}[t]
\centering \includegraphics[width=8cm]{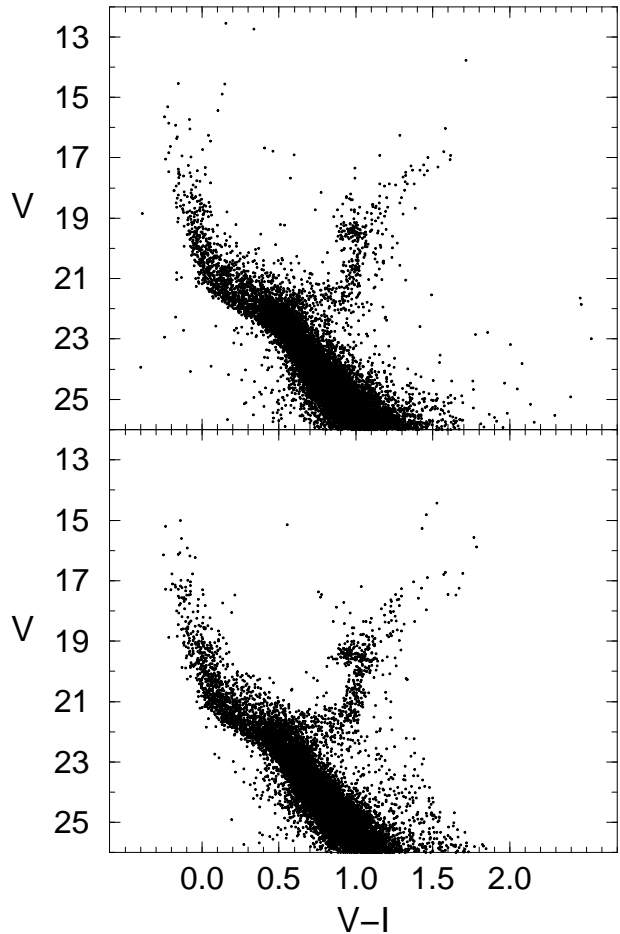}
\caption{Top panel: CMD for the selected field stars. Bottom panel:
  Best synthetic CMD.}
\label{cmds_sim_trap} 
\end{figure}

The resulting CMD, containing about 16,000 stars, is shown in the top
panel of Fig. \ref{cmds_sim_trap}.
In order to quantify the corresponding SFH we compared the observed
CMD with an extensive set of simulated histories. In such simulations,
the model behavior is changed according to the set of initial
parameters, namely the star formation law and rate SFR(t), the
chemical enrichment law Z(t), the initial mass function IMF, the
fraction of binary stars, the reddening and distance modulus. The
comparison between the observed CMD and the model CMDs is done through
minimization of the Poissonian $\chi^2$ in the number of stars in a
set of regions ($0.1\times0.1$ mag large) of the CMD
(cfr. \citealt{cignoni2010b}). In addition, following the approach
described e.g. by \cite{greggio1998}, particular attention has been
paid to reproduce the most reliable and evident evolutionary phases,
such as the blue loops, the subgiant and the red giant branches, the
main sequence Turn-Offs. To reduce computational time the SFR is
parametrized as a linear combination of fuzzy isochrones with variable
duration (partial CMDs). The final uncertainties on the SFH are
obtained with a bootstrap technique (see \citealt{cignoni2009} for
details).

A morphological comparison between observed and synthetic CMDs allows
to reduce the parameter space. We simulated models with the following
metallicity Z(t) (see e.g. \citealt{noel2009}): $Z=0.004$ for stars
younger than 2 Gyr, $Z=0.002$ for stars with age between 2 and 5 Gyr,
$Z=0.001$ for stars older than 5 Gyr. The adopted IMF has Salpeter's
exponent. A 30\% fraction of binaries is assumed. Finally, our
synthetic population is corrected for a distance modulus
${(m-M)}_{0}=18.9$ and a galactic reddening $E(B-V)=0.08$.

The first noticeable result is related to the observed UMS morphology,
whose large spread in color is not accounted for by our models either
with age or binaries or photometric errors. Even if a differential
reddening of about 0.1 mag was effective to solve this discrepancy,
the main sequence blue edge in the range $21<V<23$ would be still
redder than our models. We suggest that the youngest populations in
the field suffer both from differential reddening (which varies by
about 0.1 mag) \emph{and} from a distance modulus spread of about 0.1
mag.

Concerning intermediate to old generations ($>2$ Gyr), the compactness
of the red clump and the thinness of the red giant branch rule out any
hypothesis of differential reddening. The magnitude dispersion of the
sub-giant branch is consistent with a distance spread of at least 0.2
mag (${(m-M)}_{0}=18.8-19.00$), in good agreement with results
obtained for intermediate age star clusters (\citealt{glatt2008}). A
suggestive scenario may be summarized as follows: 1) the ``old'' field
belongs to an extended halo, evenly distributed around the average
distance ${(m-M)}_{0}=18.9$ and marginally affected by differential
reddening; 2) the ``young'' field is localized on the farther side of
the SMC and suffers differential reddening.

\begin{figure*}[t!]
\centering \includegraphics[width=12cm]{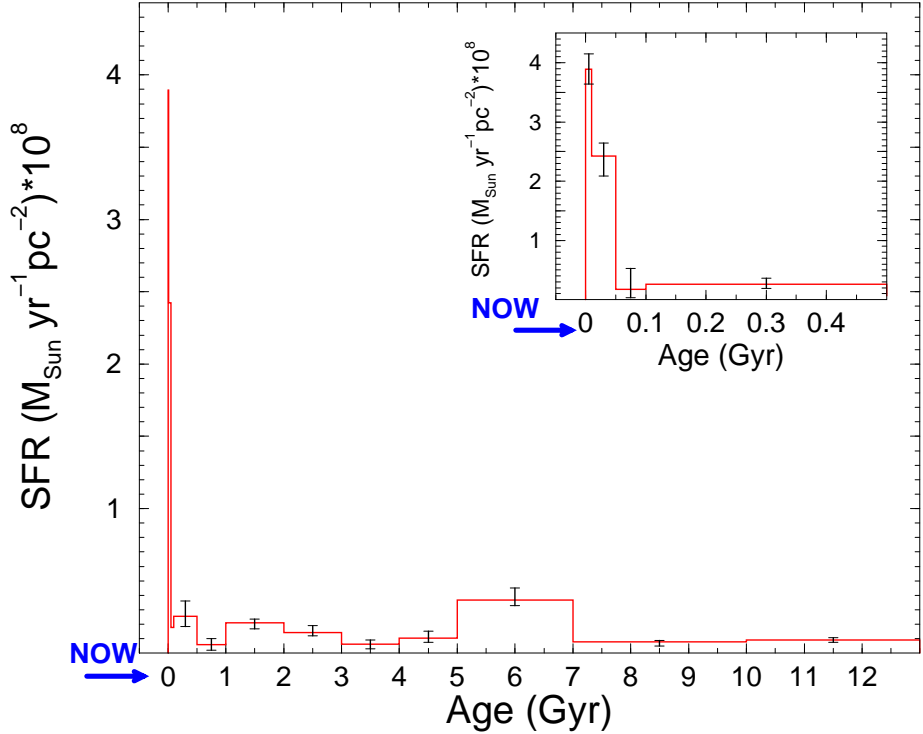}
\caption{SFH of the SMC field around NGC~346. For sake of clarity the
  most recent 500 Myr are zoomed-in in the upper right onset. The
  uncertainty on the SFR is also shown for each age bin. }
\label{sfh_rec} 
\end{figure*}

Using these assumptions on distance and reddening, we have proceeded
to recover the best combinations of partial CMDs leading to the
minimum $\chi^2$. The resulting CMD and the corresponding SFH are
shown in the bottom panel of Figure \ref{cmds_sim_trap} and in Figure
\ref{sfh_rec}, respectively. The larger errors in the recent activity
are mainly due to differential reddening. In terms of star-counts: 1)
upper and the lower main sequence (down to $V\approx\,25$) are well
reproduced; 2) the number of red clump stars is always slightly
over-predicted, by about $30\%$; 3) Blue loop stars are always
under-predicted, by $\sim\,20-30\%$.

The field SFR is increasing from 13 Gyr ago up to now (see
Fig. \ref{sfh_rec}). We estimate the average rate density in the last 100 Myr
to be $1.4\times 10^{-8}\,M_{\odot}\,yr^{-1}\,pc^{-2}$ ($2.5\times
10^{-8}\,M_{\odot}\,yr^{-1}\,pc^{-2}$ when extrapolated using a
Salpeter IMF down to $0.1\,M_{\odot}$) and the average rate density
over the whole 13 Gyr lifetime to be $1.5\times
10^{-9}\,M_{\odot}\,yr^{-1}\,pc^{-2}$ ($2.7\times
10^{-9}\,M_{\odot}\,yr^{-1}\,pc^{-2}$ when extrapolated). However,
about 60\% ($38000\,M_{\odot}$) of the stellar mass was assembled in
the earliest 8 Gyr, with 36\% ($21000\,M_{\odot}$) formed between 7
and 5 Gyr ago. For comparison with another SMC star forming region,
NGC~602, in the period between 3 - 13 Gyr ago the field around NGC~346
produced at least six times more mass per $pc^2$ than the field around
NGC~602 (\citealt{cignoni2009}).
Finally, it is also comforting to note that no significant SFR peak in
the more recent period 5-4 Gyr ago is found, showing that the
contamination from BS~90 is actually minimal.

\section{Discussion and Conclusions}

The analysis of the CMDs and of the spatial distribution of the
stellar populations in the region of NGC~346 leads to interesting
results, both for the clustered and the diffuse components.

\subsubsubsection{Diffuse component}

For the diffuse component we find that the majority (60\%) of the
mass astration occurred earlier than 5 Gyr ago, with a peak between 5
and 7 Gyr ago. This is at variance with the suggestion by
\cite{harris2004} of a long period of almost no activity from 3 to 8.4
Gyrs ago, but in agreement with the SFH of other SMC regions derived
by other authors (e.g. \citealt{dolphin2001}, \citealt{mccumber2005},
\citealt{cignoni2009}, \citealt{noel2009}, \citealt{sabbi2009}). We
recall that our photometry and those of all the latter authors reach
the oldest MS turn-off and therefore a look-back time of 13 Gyr, while
the shallower data by Harris \& Zaritsky do not allow to properly
describe epochs earlier than a few Gyr.

Over the SMC lifetime the average rate density in the field
surrounding NGC~346 is $1.5\times
10^{-9}\,M_{\odot}\,yr^{-1}\,pc^{-2}$. This value is in agreement with
those derived by \cite{noel2009} for several SMC regions. Its SFH is
qualitatively similar to that around the other star forming region we
studied with HST/ACS data, NGC~602 (\citealt{cignoni2009}). However,
the activity in the NGC~346 region is significantly higher (by a
factor of 6) than in NGC~602. These results are not surprising given
the much closer proximity of NGC~346 to the SMC center. The rate of
star formation in the NGC~346 region (and in NGC~602 as well) appears
to have increased in the last tens of Myr from a relatively lower and
steady regime. We must remind, however, that the SFH recovered at very
old epochs is averaged over timescales of Gyr and short episodes of
higher activity may remain undetected.

In the last 100 Myr the average star formation rate density in the
field is about $1.4\times 10^{-8}\,M_{\odot}\,yr^{-1}\,pc^{-2}$, a
factor of 2-10 higher than in the solar neighborhood
(cfr. \citealt{timmes1995}), two orders of magnitude higher than in
nearby late-type dwarfs, and similar to the quietest cases of Blue
Compact Dwarfs (see \citealt{tolstoy2009} and references therein). In
this respect, the SMC region around NGC~346 appears as currently quite
active, but we should bear in mind that its current rate might be
somewhat overestimated by the possible presence of very young stars
member of the actual star forming region itself. For comparison, the
peak rate in SC-1 is $2\times 10^{-5}\,M_{\odot}\,yr^{-1}\,pc^{-2}$,
three orders of magnitude higher than the field average value.

\subsubsubsection{Clustered component}

For the clustered component, we have found that the bulk of the stars
in Group I sub-clusters are well consistent with a major star forming
episode started about 6 Myr ago and lasted about 3 Myr. After that, their
star formation activity has proceeded at a lower rate. As suggested by
\cite{palla2002} for the stellar group Taurus-Auriga, massive stars,
dispersing the gas that would have been part of the new generations,
may be the culprit for such ``deceleration'' in the most recent star
forming activity.

Our models do not fully explain the large color spread shown by PMS
stars and the large observed number of PMS stars.  The PMS shows
indeed a redward broadening which is not reproduced by either our
youngest isochrones or photometric scatter. \cite{hennekemper2008}
suggested that a combination of differential reddening, variability
and binarity may account for the observed spread. We consider variable
reddening affecting individual PMS stars a more likely explanation
(also taking into account that our PMS evolutionary tracks do not
include circumstellar reddening). Moreover, our models seem to
underestimate the observed PMS counts by about 40\% (possibly and in
part because our PMS tracks do not cover masses smaller than
$0.45\,M_{\odot}$). We speculate that a very young generation of
stars, so young not to have had time to assemble more massive stars,
could account both for the observed excess of PMS stars and for the
intrinsic redness of such stars; otherwise, the IMF must be steeper
than Salpeter.

While the origin of these discrepancies is still unclear, the
comparison with CMDs of stars in different locations provides clues to
complete the puzzle.

Low mass PMS stars dominate the star counts among Group II
sub-clusters, showing a ratio PMS/UMS stars that is at least three
times higher than in Group I. Such a population should have a MS
counterpart, which is however not visible.  Among all the sub-clusters
in NGC~346, those in Group II host the reddest PMS stars. Only a
strong and differential reddening correction could reconcile the Group
II PMS with the SC-1 PMS, but it would pay the price of exacerbating
the lack of UMS stars. Vice versa, an age effect, with the Group II
sub-clusters experiencing now their first episode of star formation,
may explain both the PMS redness and the higher PMS/UMS ratio. If the
NE-Spur, where the Group II sub-clusters are located, is indeed the
youngest region of NGC~346, a fraction of stars may have managed to
retain the circumstellar material and appear redder. Moreover, the
youth of the system and the peculiar distribution of NE-Spur stars,
preferentially located along filaments, may determine the paucity of
UMS. In this case the peak gas densities in the filaments would be
insufficient to produce massive stars, such as those detected in
SC-1. As a consequence, the ratio of low-mass to massive stars is
higher where the density of massive stars is lower.

This trend is also noted by \cite{panagia2000} in the field of
SN1987A. It indicates that star formation processes for different
ranges of stellar masses are rather different and/or require different
initial conditions. An interesting corollary may be that, if the
youngest objects have a filamentary distribution (where presumably the
gas density was higher), \emph{a critical density threshold exists
  below which the star formation is suppressed}. The reason for this
is probably related to the pristine conditions in the parental
molecular clouds, like temperature or turbulence.

As an alternative to environmental effects, the lack of UMS stars may
reflect the build up timescales of different masses. More massive
stars may form later than low-mass stars because they need more time
to collect enough material to start the formation. In this case the
Group II sub-clusters are simply too young to have produced massive
stars.

The analysis of \cite{hennekemper2008} and \cite{gouli2008} provide an
independent support to the age hypothesis. Fitting Seiss isochrones,
\cite{hennekemper2008} derived for the NE-Spur sub-clusters ages younger
than 2.5 Myr, while \cite{gouli2008} concluded that the star formation
there was recently triggered (see figure 1 in \citealt{gouli2008}) by
the massive progenitor of SNR B005-7724. We notice however that
\cite{naze2002} argued that the SNR should be located in front of
the luminous blue variable HD-5980 and not really within the NGC~346
region.

From the point of view of the gas, the NE-Spur is a natural place to
find new stellar generations. Spectroscopic observations of CO prove
the existence of reservoirs of cold molecular gas in the spatial
region around SC-14 and SC-15 (see \citealt{rubio2000}) while the
distribution of the radiation field at 160 nm is clearly confined to
the Bar, and it is not detected in the Spur (except in SC-13).  In
other words, the extreme NE-Spur contains residual reservoirs of
molecular gas that may be fueling the star formation.

The Group III sub-clusters exhibit: 1) the same anomalous ratio PMS/UMS
stars found in Group II (hence at variance with Group I); 2) a color
spread and redness among LPMS that is perfectly in line with what is
found in Group I (hence at variance with Group II); 3) a color spread
among UMS that is at variance both with Group I and with Group II
members.

 This suggests a picture where the Group III sub-clusters are in an
 \emph{intermediate state} between those in Group II and those in
 Group I. If the underlying parameter is the age, these sub-clusters
 are sufficiently old that low mass PMS stars have already completed
 their accretion phase (hence their color dispersion is consistent
 with the ``evolved'' cluster SC-1), but at the same time so young
 that intermediate and massive stars are still approaching the UMS. As
 for the NE-Spur, another key of interpretation is the available gas
 out of which the presently observable stars were assembled. In this
 picture, the gas density in the NW-Bar was such that intermediate and
 massive stars were formed more slowly than in the SE-Spur.

These findings appear to support the view that low-mass stars form
more ``easily'' than massive ones either because they need less gas
density or lower rates of accretion. Observational support for
this was already presented e.g. by \cite{ruppert2009}. Likewise, the
Spitzer detection (\citealt{simon2007}) of proto-OB stars in Group II
and Group III sub-clusters may explain the apparent lack of UMS stars
in terms of age.

As comprehensively reviewed by \cite{zinnecker2007} no consensus has
been reached yet on which is the most likely process for massive star
formation: monolithic collapse in isolated cores, competitive
accretion in a protocluster environment or stellar collision and
mergers in very dense systems. NGC~346 seems to favor one of the
latter two (or both) but we should wait for further high-resolution
observations at longer wavelengths (e.g. with WFC3 on board of HST,
but also ALMA and JWST, and eventually with ELTs) to get a better
insight in its SF process based also on its still embedded, not
visible, youngest objects.

In the conditions described above an assessment of the IMF in the
NGC~346 region is risky. There is no doubt that in several
sub-clusters the number of existing massive stars is definitely lower
than predicted by a Salpeter IMF. \cite{sabbi2008} already pointed out
that in the region massive stars are underrepresented.  We speculate
that the paucity of UMS in NGC~346 has a double origin: the maximum
mass populating Group II sub-clusters is a consequence of the
radiation feedback, while the maximum mass populating the PMS
sub-clusters is mainly a matter of youth. From a general point of
view, our result goes in the same direction of a relation between the
mass of the most-massive star in the cluster and the mass of its
parent star cluster as suggested by \cite{weidner2006}. On the other
hand, if indeed clusters form in an ordered fashion producing first
low mass stars and then proceeding to assemble more massive stars till
a maximum mass is born whose feedback halts the collapse, how is it
possible to explain the existence of sub-clusters hosting only PMS
stars? We expect Alma, Herschel and JWST to provide the necessary
clues.

\section*{Acknowledgments}
MC and MT acknowledge financial support through contracts
ASI-INAF-I/016/07/0 and PRIN-MIUR-2007JJC53X-001. Partial support for
U.S. research in program GO10248 was provided by NASA through a grant
from the Space Telescope Science Institute, which is operated by the
Association of Universities for Research in Astronomy, Inc., under
NASA contract NAS 5-26555.


\end{document}